# Properties of $Al_2O_3$ thin films deposited on 4H-SiC by reactive ion sputtering

P. Fiorenza[1)], M. Vivona[2)], S. Di Franco[1)], E. Smecca[1)], S. Sanzaro [1,3)], A. Alberti [1)], M. Saggio [4)], F. Roccaforte[1)]
[1)] CNR-IMM, Strada VIII, n. 5 - Zona Industriale, Catania, Italy
[2)] Optoelectronics Research Centre (ORC), University of Southampton B53, SO17 1BJ Southampton, UK
[3)] Dipartimento di Fisica e Scienze della Terra, Università degli Studi di Messina, Italy
[4)] STMicroeletronics, Stradale Primosole 50, Catania, Italy

## ABSTRACT

In this work, the electrical properties of $Al_2O_3$ films deposited by reactive ion sputtering were investigated by means of morphological, chemical and electrical characterizations. This insulating layer suffers of an electron trapping that is mitigated after the rapid thermal annealing (RTA). The RTA improved also the permittivity (up to $6\varepsilon_0$), although the negative fixed charge remains in the order of $10^{12}cm^{-2}$. However, the temperature dependent electrical investigation of the MOS capacitors demonstrates that the room temperature Fowler-Nordheim electron barrier height of 2.37 eV lies between the values expected for $SiO_2$/4H-SiC and $Al_2O_3$/4H-SiC systems.

## INTRODUCTION

For several years, silicon carbide (4H-SiC) has been studied considering its potential applications for high power devices. Owing to this continuous impulse of the scientific community, the achievements in material quality and processing maturity led finally to the commercialization of several SiC power devices (i.e., diodes and MOSFETs) and modules [1]. Hence, nowadays novel scientific and technological problems are under consideration to improve or even to go beyond the currently available technologies [2]. In this context, while $SiO_2$ is commonly used for 4H-SiC metal oxide semiconductor field effect transistors (MOSFETs), a future issue will be related to the study and optimization of novel gate dielectric materials. In particular, the introduction of a dielectric with a high permittivity (*high-k*) is expected to improve the device performance and reliability in high-voltage applications [3]. In fact, the gate insulator in 4H-SiC MOSFETs must withstand a high electric field. In particular, in blocking configuration, the electric field in the gate insulator ($E_{ins}$) is related to the electric field in the semiconductor ($E_s$) by the Gauss' law, $E_{ins} = (\kappa_s/\kappa_{ins})E_s$, where $\kappa_{ins}$ and $\kappa_s$ are the insulator and semiconductor permittivity values. Hence, to mitigate the stress in the gate insulator, its permittivity has to be close to the SiC permittivity.

However, in the choice of an alternative gate dielectric for 4H-SiC devices, the permittivity is not the unique factor to be considered. In fact, the band gap, band alignment and thermal stability are also key parameters. Namely, a high permittivity should be preferably combined with a high barrier height, in order to reduce the leakage current [3].

Because of its interesting properties, such as a high permittivity ($\kappa$~7-9), thermal stability up to 1000°C, and reasonable conduction band offset with respect to 4H-SiC (> 1.5 eV as shown in Fig. 1), aluminum oxide ($Al_2O_3$) has already attracted the interest as alternative dielectric in SiC-based devices [4,5]. Typically, the $Al_2O_3$ films, deposited by chemical vapor deposition (CVD) or atomic layer deposition (ALD), have shown a high trapped charge density (and flat band voltage shifts) and significant hysteresis during cyclic C-V measurements.

In this work, the electrical properties of $Al_2O_3$ films deposited by reactive ion sputtering were investigated by means of MOS capacitors. The capacitance vs voltage characteristics collected on the different samples revealed the beneficial effect of a rapid thermal treatment that improved the permittivity and reduced the electron trapping in the insulating stack. The hysteresis is strongly reduced after the thermal treatment with respect to the as-deposited sample. Furthermore, temperature-dependent current vs voltage characteristics allowed to investigate the conduction mechanisms through the insulating stack, revealing a nearly ideal effective tunneling mechanism with an electron barrier height that lies between the ideal values for $SiO_2$ and $Al_2O_3$ layers. However, the low dielectric breakdown field is still a concern that deserves to be improved. This deposition method is rather cost-effective compared to CVD and ALD and does not present safety concerns instead related to the use of metal-organic precursors.

## EXPERIMENTAL

A schematic of the fabricated MOS capacitors and of the theoretical band alignment in the investigated 4H-SiC/$SiO_2$/$Al_2O_3$ system is reported in Fig. 1. The MOS capacitors were fabricated on n-type epitaxial 4H-SiC layers with a donor concentration of $1\times10^{16}$ cm$^{-3}$, grown onto heavily doped substrates. First of all, a thermal oxidation in dry $O_2$ was carried out at 1150°C in order to grow a 6nm thick interfacial $SiO_2$ barrier layer. This thin oxide was introduced to increase the oxide/semiconductor barrier height (up to 2.7 eV). Indeed, in the absence of this thin $SiO_2$ barrier layer it was not possible to perform a reliable electrical analysis, due to the high leakage current

Thereafter, a large area Ohmic contact side was formed on the sample back by Ni-deposition followed by a rapid thermal annealing (RTA) at 950°C [6], prior to $Al_2O_3$ deposition on the front. The thin $Al_2O_3$ film was deposited by reactive ion sputtering from an Al target in oxygen ambient at low deposition rate. The reactive ion sputtering was carried out at base pressure of $5\times10^{-7}$mbar and at generator conditions of 20W, 274V, 50kHz. The resulting deposition rate was 0.4 nm/min. During deposition the sample was placed on a sample holder under rotation at 20 revolutions per minutes to reduce materials inhomogeneities.

The electrical characterization of the deposited films was carried by capacitance–voltage (C–V) and current–voltage (I–V) measurements at different temperatures, using a Cascade probe station and a Keysight B5105 parameter analyzer. The morphology of the as deposited and annealed insulators was investigated using a PSIA XE-150 atomic force microscope (AFM) operating in noncontact mode with highly doped silicon tips.

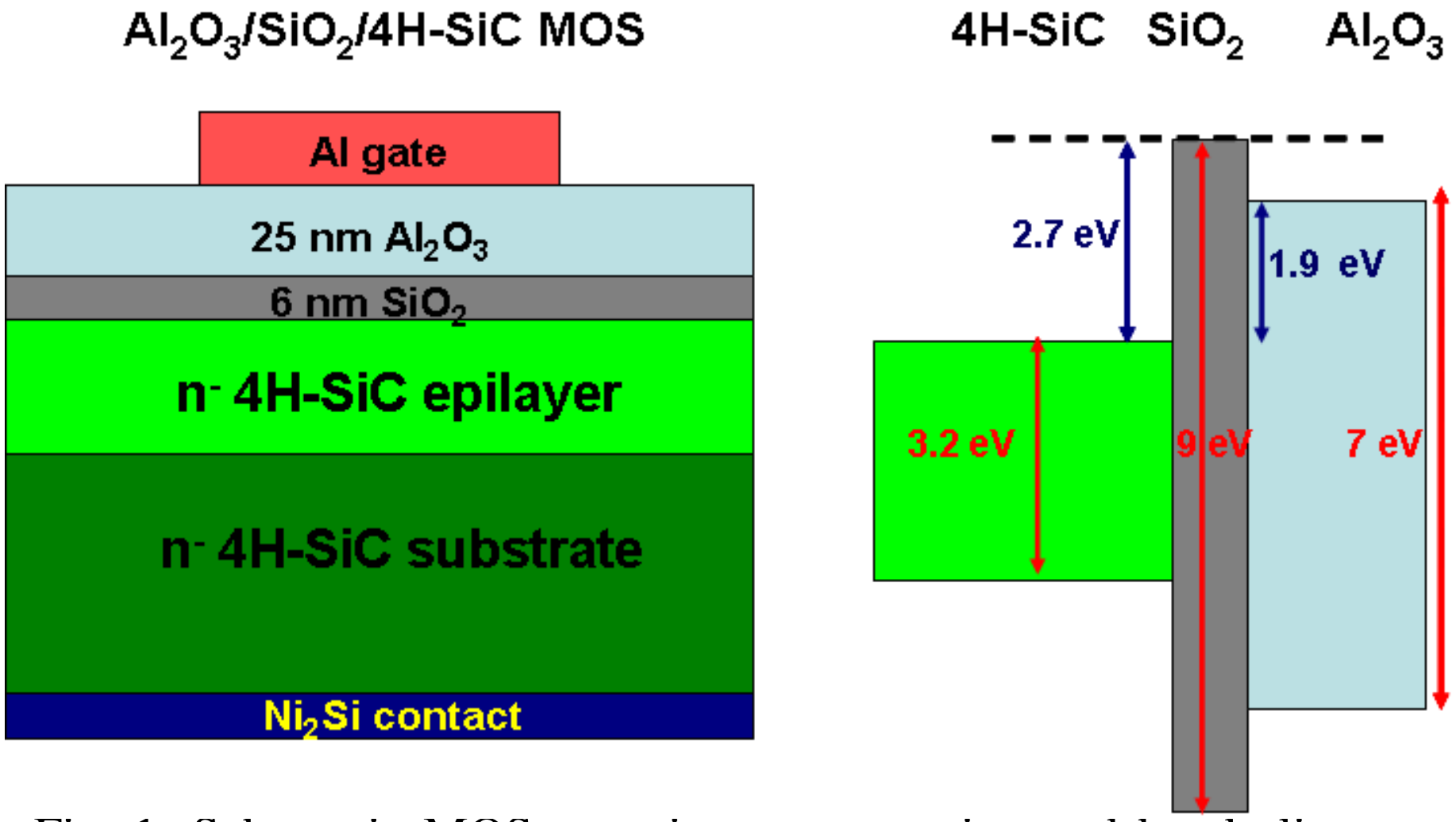

Fig. 1: Schematic MOS capacitor cross section and band alignment diagram for the 4H-SiC, $SiO_2$ and $Al_2O_3$ system.

## RESULTS AND DISCUSSION

First, the composition and surface morphology of the insulating layers were monitored.

The thin film composition was investigated with an ex-situ chemical characterization on blanket samples by XPS (Fig. 2) confirmed the formation of stoichiometric $Al_2O_3$. The thickness of this film was about 25 nm, as determined by ellipsometry measurements. Furthermore, the insulating $Al_2O_3$ films were annealed with a RTA process in $N_2$ at 800°C for 60 seconds. Finally, the film morphology was studied by employing the atomic force microscopy (Fig. 3). The final film is conformal and smooth with a root mean square (RMS) of the height distribution of 2.0 nm. The MOS capacitors were finalized with the deposition on the sample front of the gate electrode by a 100 nm Al-film deposition and photolithography processes.

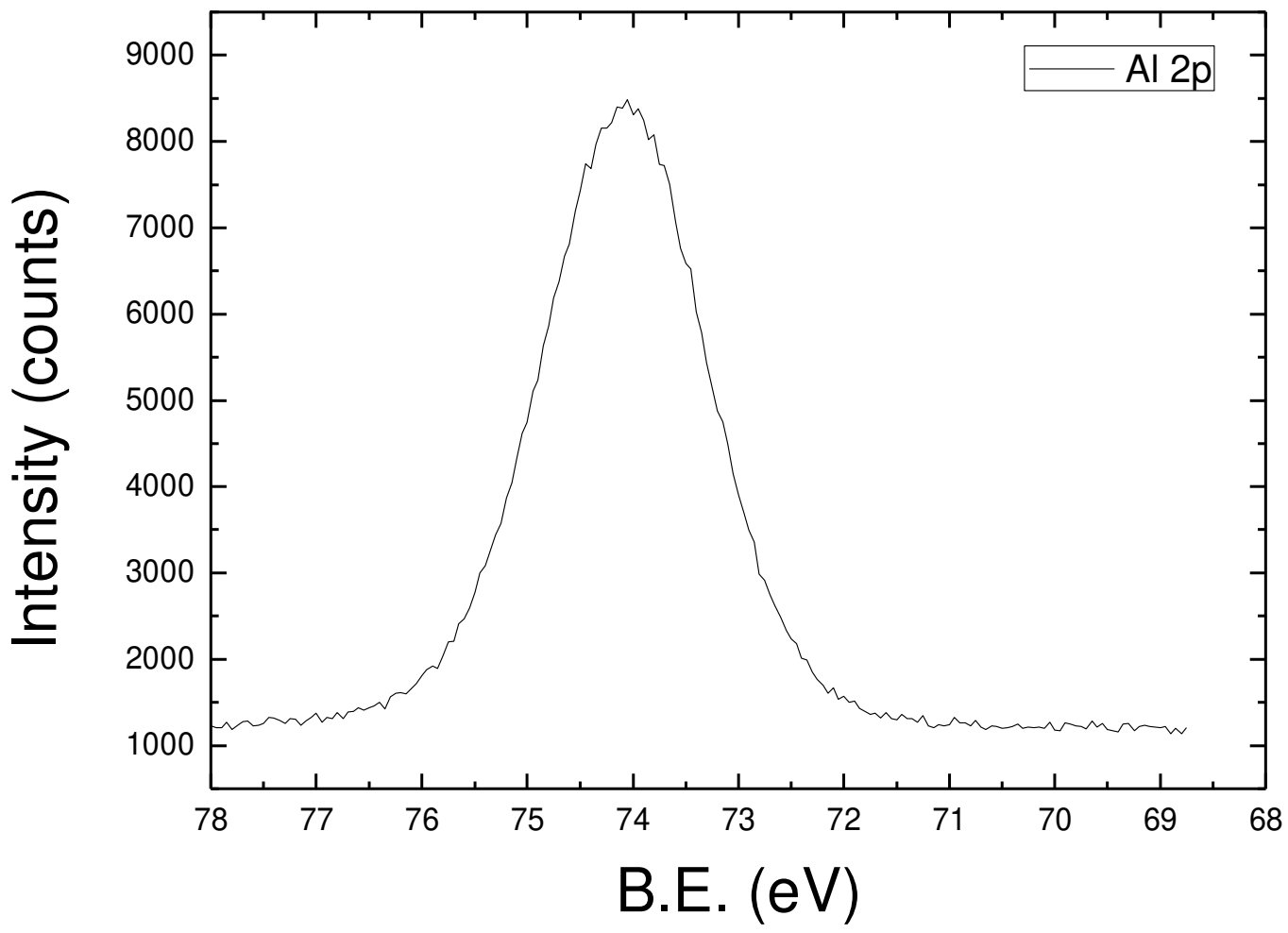

Fig. 2: XPS spectrum collected on the bare insulator surface.

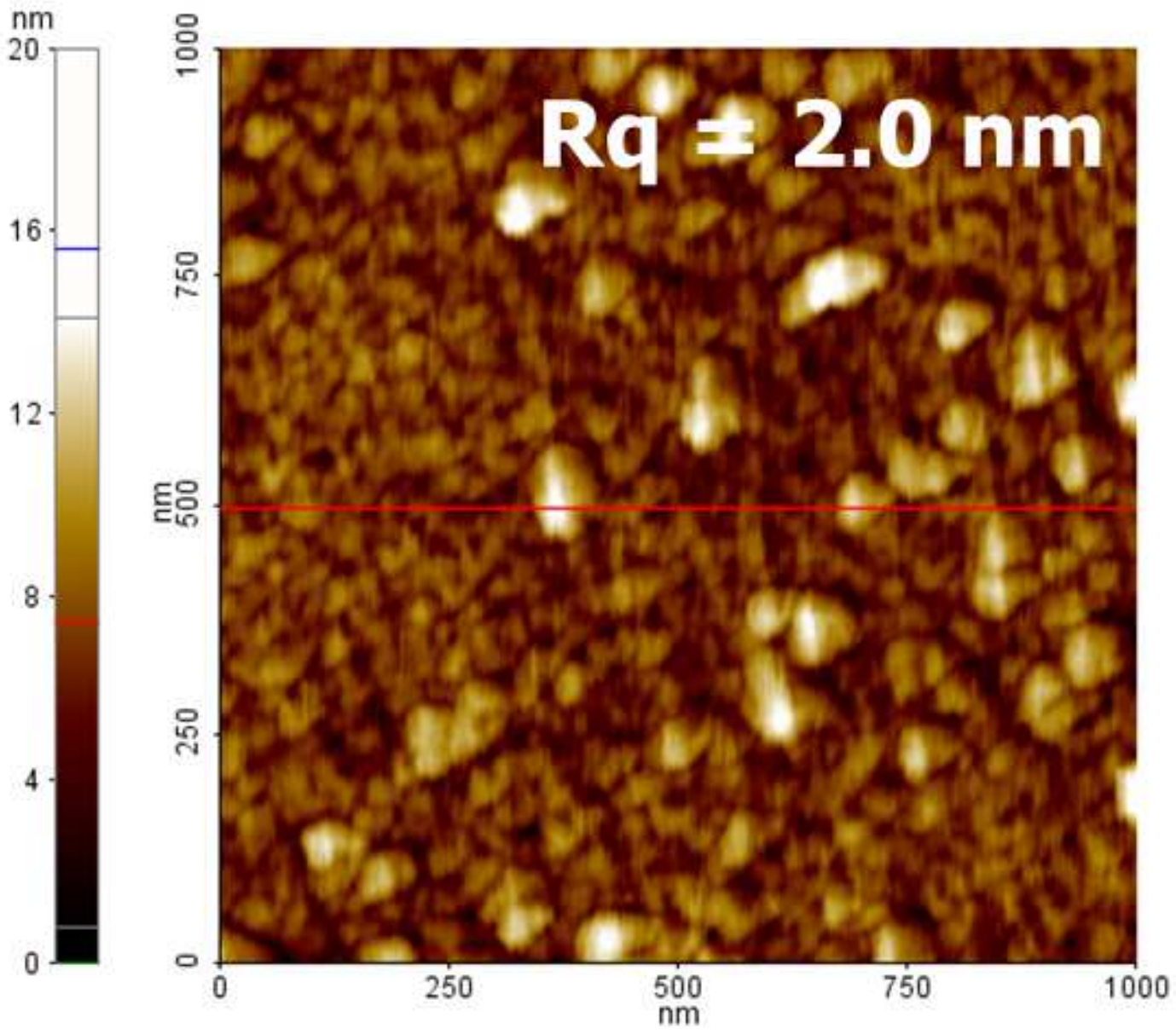

Fig. 3: AFM morphology map collected after the thermal annealing.

Thereafter, the electrical properties of the system were studied by means of I-V and C.V measurements on MOS capacitors.

Fig. 4 shows the C-V curves acquired on the MOS capacitors both before and after RTA in $N_2$. Evidently, before annealing a considerable positive shift of the flat band voltage $\Delta V_{FB}$ = +5V is observed after the first bias sweep from depletion toward accumulation. This behavior indicates the presence of a large amount of negative charges inside the oxide, i.e., in the order of $2\times10^{12}cm^{-2}$. After the first measurements, repeated measurements sweeping from depletion toward accumulation and backward, demostrate that the curves are permanently shifted toward more positive gate bias. Clearly, a stable negative charge trapping occurred during the C-V measurements. Furthermore, from the value of the accumulation capacitance $C_{ox}$ it was possible to estimate a permittivity value of 3.3 $\varepsilon_0$. The permittivity evaluation takes into account that $C_{ox}$ is the series capacitance of the two layers in the insulator stack. The low permittivity value can be attributed to lower density of the as-sputtered film. However, after the annealing at 800°C, an improvement of the C-V curves is observed, as the pristine flat band voltage shift is significantly reduced, thus indicating a drastic decrease of the net negative effective charge in the oxide layer. Furthermore, the $C_{ox}$ accumulation capacitance is increased and considering the physical thickness of the $SiO_2$ and $Al_2O_3$ layers, it was possible to determine an increase of the permittivity up to 6 $\varepsilon_0$ for the deposited $Al_2O_3$ films after the RTA. This value is fully in line with the literature values obtained with other growth methods.

The reduction of the net negative trapped charge in the oxide during the measurements obtained after RTA process is deduced by the significant reduction of the hysteresis in the C-V curves. In fact, as can be observed in Fig. 4, reporting the cyclic C-V curves acquired from depletion to accumulation and backward in the two cases, the hysteresis was almost completely suppressed in the annealed oxide; i.e., in the order of $1\times10^{11}$cm$^{-2}$. However, in literature $Al_2O_3$ thin films synthesized using different techniques (e.g., ALD) suffer of similar electron trapping phenomena, likely due to the presence of oxygen vacancies [7].

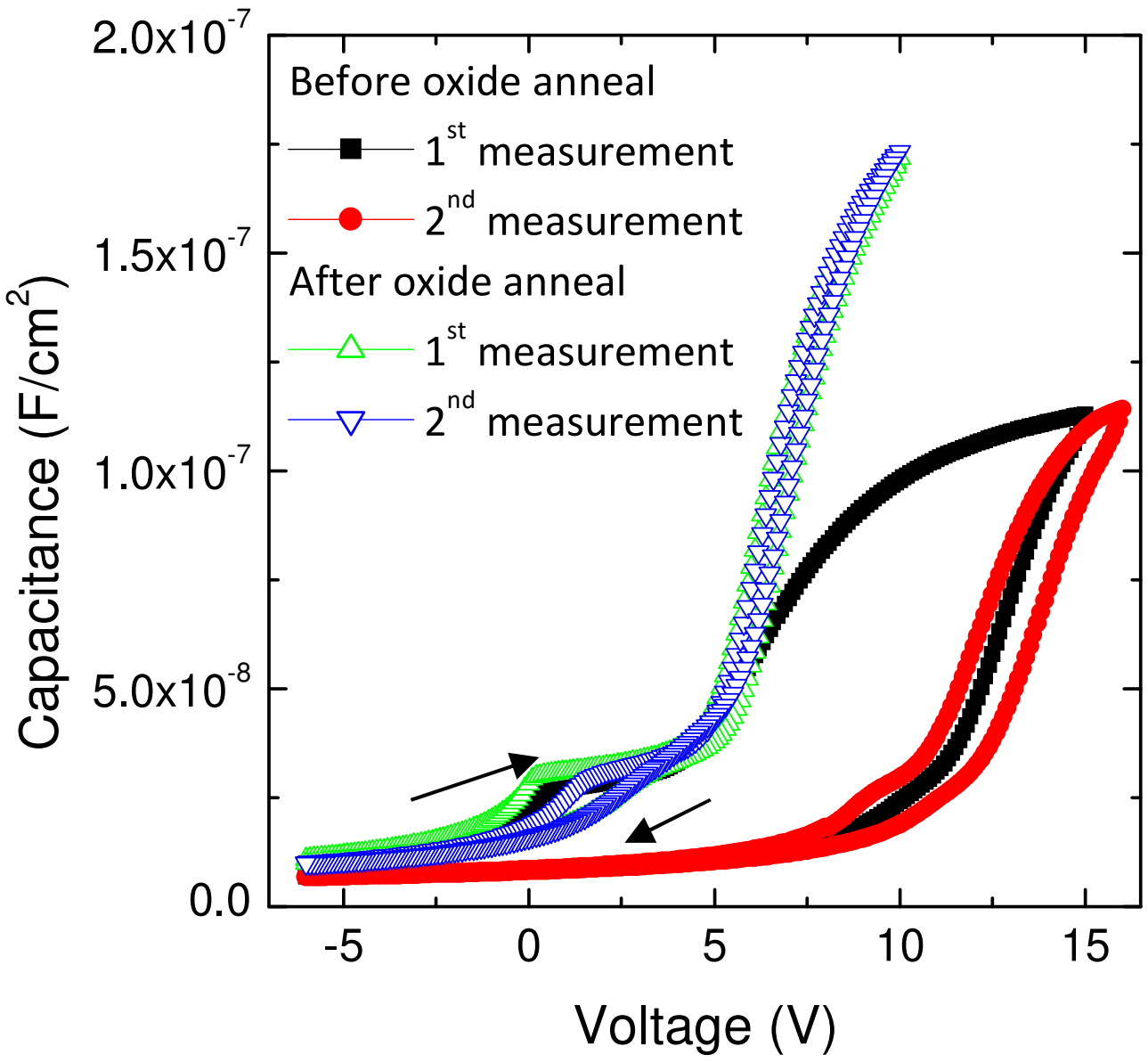

Fig. 4: C-V curves collected on $Al_2O_3/SiO_2$/4H-SiC MOS capacitors after two consecutive bias sweeps from depletion to accumulation, both for the as deposited sample and for the sample subjected to RTA at 800°C

Comparing the C-V curves on the as-deposited and RTA samples it is possible to notice that the flat band voltage ($V_{FB}$) ranges in both cases between + 6V and + 7V. The $V_{FB}$ is far from the ideal value (~0V), thus indicating the persistence of a residual negative fixed charge , i.e., in the order of $2\times10^{12}$cm$^{-2}$. Furthermore, the forward curve – from depletion to accumulation – shows a knee that is nearly absent in the reverse curve indicating the occurrence of energetically deep interface states trapping (in the order of $1\times10^{12}$cm$^{-2}$).

Another import aspect of the characterization comes from the gate dielectric breakdown statistics. Fig. 5a shows a I-V characteristic collected for a MOS capacitor. The I-V curve shows a dielectric breakdown of the oxide occurring at approximately 14 V, corresponding to an electric field of 4.5 MV/cm. By performing statistical measurements on capacitors, we are able to apply the Weibull formalism to the experimental breakdown data. In particular, the Weibull distribution is given by [8]

$$F(x) = 1 - exp\left[-(x/\alpha)^{\beta}\right].$$

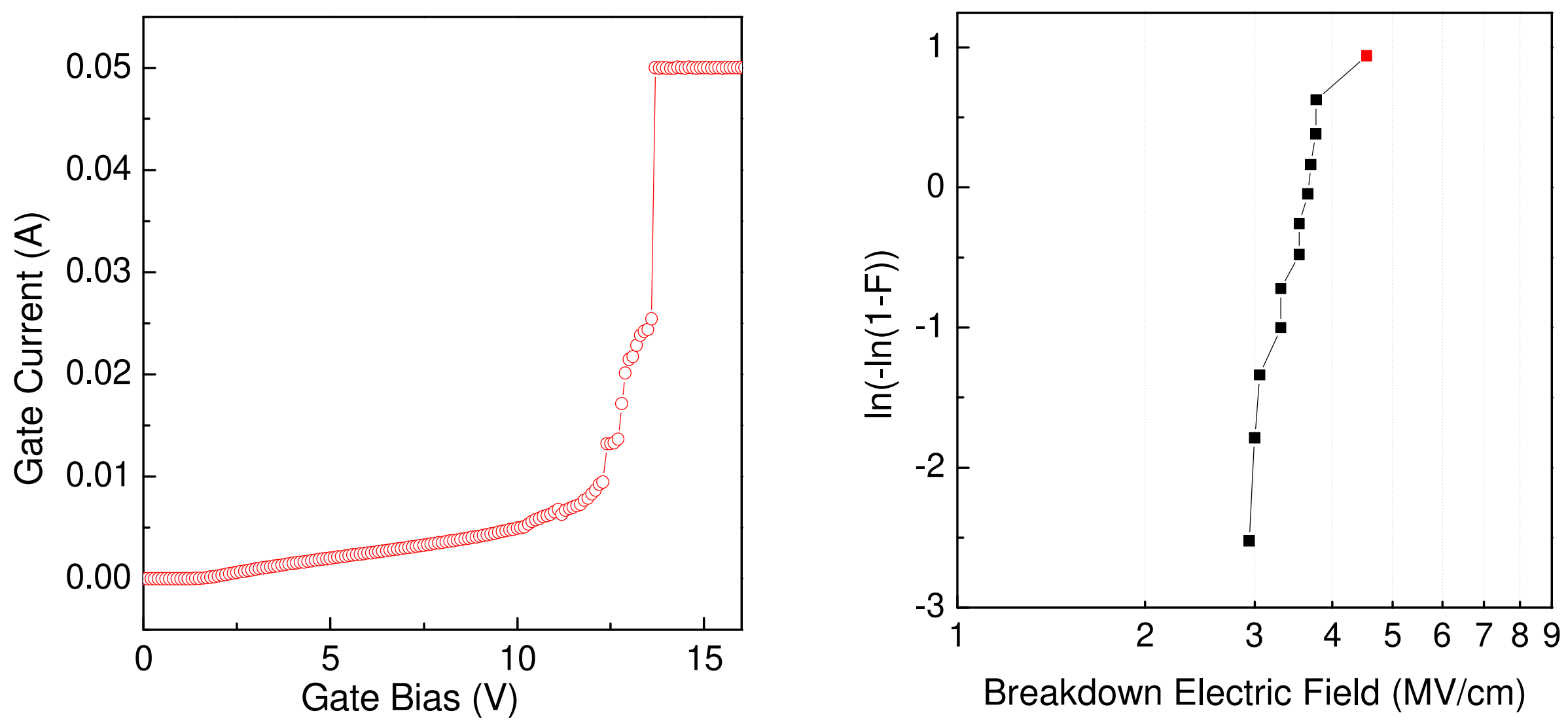

where F is the cumulative failure probability (representing in our case the ratio between the failed and the investigated MOS $N_f /N_i$), x is the electric field that induces the BD phenomena, α is the characteristic lifetime of the dielectric, and β is called the Weibull slope. In particular, the ln[-ln(-F)] of the MOS capacitors is plotted versus the natural logarithm of the BD electric field. As a consequence, the more robust is the insulator the larger the β value is and the steeper the Weibull plot is. As can be seen in Fig. 5b the BD occurred mostly between 3 and 5 MV/cm on the investigated MOS capacitors. However, even if the BD statistic is centered in a narrow electric field range (3-5 MV/cm), it is far away from the ideal value (~10MV/cm).

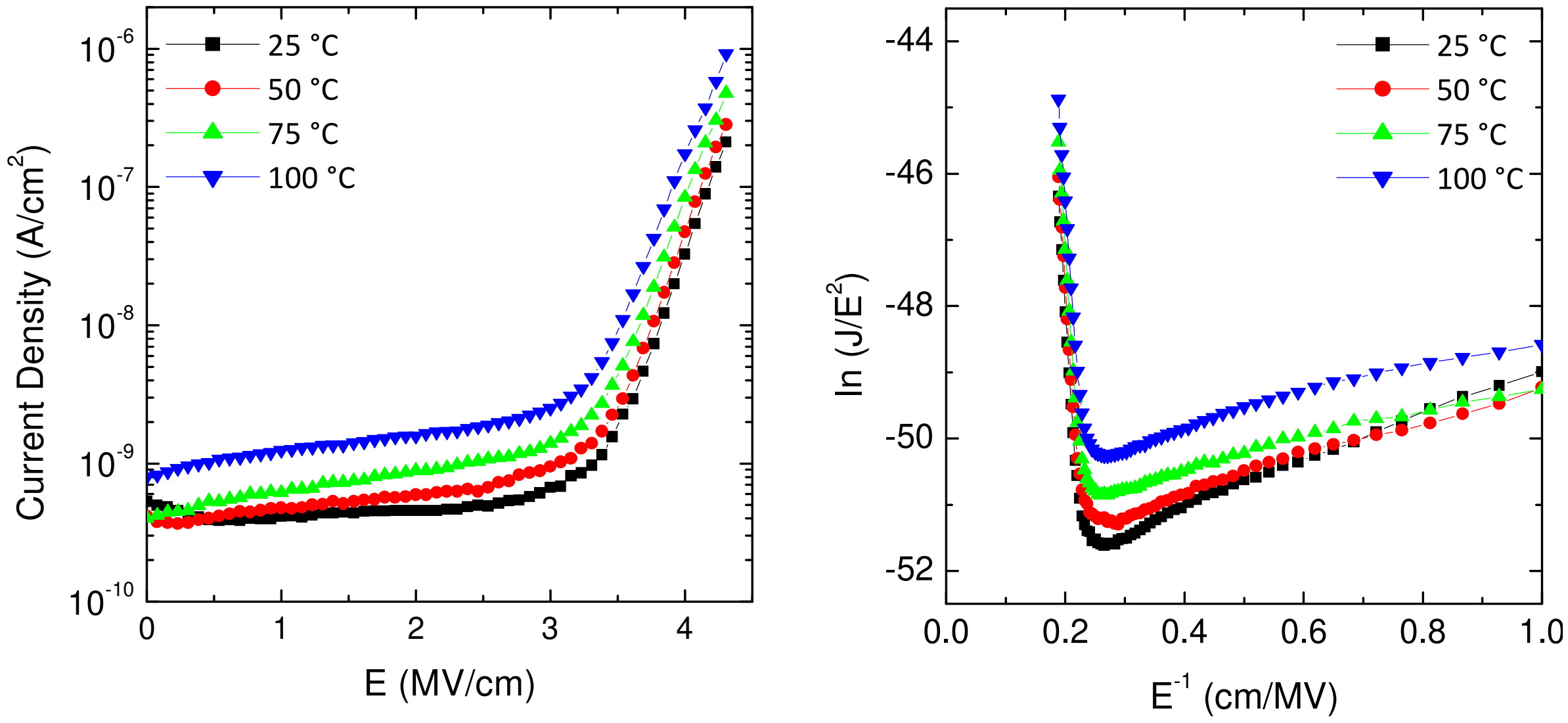

The insulating properties were also studied by monitoring the temperature dependence of the I-V characteristics of MOS capacitors, from room temperature up to 100 °C.

Once a MOS device is stressed and the electron trapping occurred, it becomes stable and it is possible to compare the I-V collected at different temperature. The I-V characteristics were analyzed using the Fowler-Nordheim (FN) formalism and the Lenzlinger-Snow (LS) equation:

$$ln\left(J/E^2\right) = ln\left(\frac{q^3\left(m_{SiC}/m_{ox}\right)}{8\pi h\Phi_B}\right) - \frac{8\pi\sqrt{2m_{ox}\Phi_B^3}}{3qh}\frac{1}{E}$$

where J is the current density, E is the electric field across the oxide, $m_{SiC}$ and $m_{ox}$ are the effective electron masses in the SiC substrate and in the insulator respectively, q is the electron charge, h is the Plank constant and $\Phi_B$ is the effective tunneling barrier height for electrons [9]. By the fits (not shown) in the linear region of the "FN plots" $ln(J/E^2)$ vs 1/E shown in Fig. 6b, it was possible to determine the values of $\Phi_B$ for electrons as a function of the temperature.

As shown in Fig. 7 the FN tunneling can be used to describe the current conduction mechanisms of the system. In fact, ideally the FN tunneling possesses a weak temperature dependence, related to the shrinking of the SiC and insulator band gaps with increasing temperature [9]. On the other hand, it has been demonstrated that thermally grown $SiO_2$ layers onto 4H-SiC may contain residual carbon atom content that can affect the insulating properties of the gate oxide [10]. In defective insulators Poole-Frenkel (PF) emission can often rule the conduction mechanism. The ideality of the tunneling mechanism can be investigated looking at the variation of the fitting barrier height value and in particular looking at its slope $d\Phi_B/dT$. In literature, some thermal oxides grown onto 4H-SiC shown an experimental $d\Phi_B/dT$ slope (–7.6 meV/°C) one order of magnitude higher than the theoretical (–0.7 meV/°C) one and the experimental $d\Phi_B/dT$ slope (–0.98 meV/°C) of the deposited oxide, respectively [9,10].

Fig. 7 shows how both our experimental results, literature data and the ideal values of the electron barrier height $\Phi_B$ vary with a temperature increasing temperature from 0 °C up to 250 °C for the $SiO_2$ and the $Al_2O_3$ cases. The ideal $SiO_2$ FN case possesses a barrier height $\Phi_B$ = 2.7 eV while ideal $Al_2O_3$ FN case possesses a barrier height $\Phi_B$ = 1.9 eV. In both cases the barrier height decreases with a slope of the $d\Phi_B/dT$ = – 0.7 meV/°C [11] with increasing temperature.

On the other hand, the experimental effective barrier height for the $SiO_2/Al_2O_3$ stack is $\Phi_B$ = 2.37 eV at room temperature, i.e., larger of the ideal value expected for $Al_2O_3$/4H-SiC interface and smaller than the $SiO_2$/4H-SiC interface. This result suggests that the FN formalism averaged the effective double tunneling occurring on the $SiO_2/Al_2O_3$ stack. However, an indication of the good quality of the insulating stack is obtained considering that the experimental slope of the $d\Phi_B/dT$ is – 1.07 meV/°C is only slightly larger than the ideal value of – 0.7 meV/°C. Indeed, the experimental results on the $SiO_2/Al_2O_3$ stack are improved compared to some thermally grown $SiO_2$ layers [10].

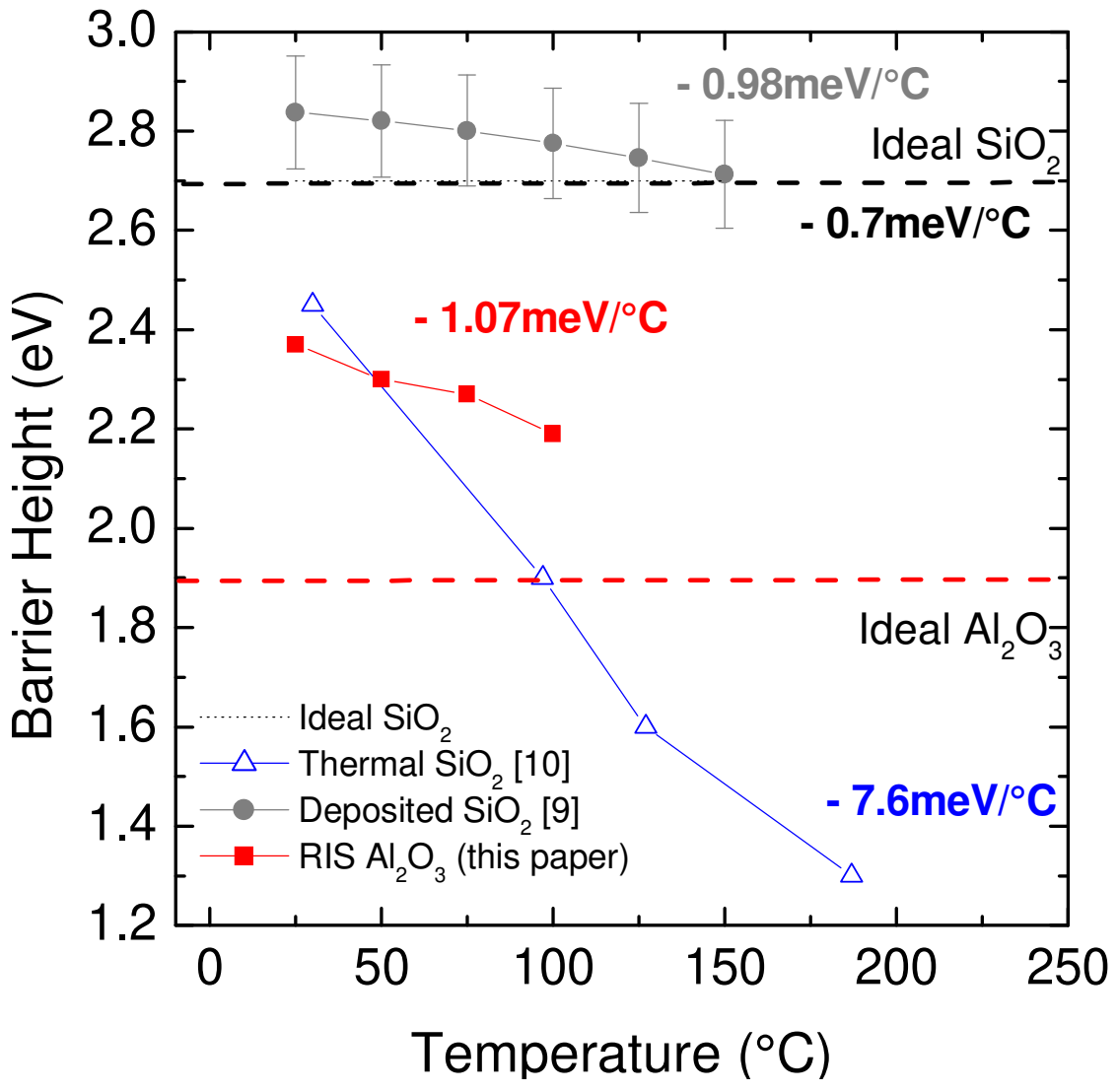

Fig. 7: Comparison between the experimental electron barrier height for the $SiO_2/Al_2O_3$ stack (this paper) and literature data on $SiO_2$, and the theoretical behavior.

## CONCLUSION

In this paper, we have investigated the morphological, chemical and temperature dependent electrical properties of $Al_2O_3/SiO_2$/4H-SiC system in MOS capacitors. The $Al_2O_3$ films were grown by reactive ion sputtering onto a thin thermal $SiO_2$ layer. The insulating layer suffers from an electron trapping that is mitigated after the RTA. The RTA improved also the permittivity, without significantly reducing the negative fixed charge. The temperature dependent electrical investigation of the MOS capacitors resulted in a Fowler-Nordheim barrier height of 2.37eV. The Fowler-Nordheim-like tunneling has been proved looking at the temperature coefficient of the electron barrier height ($d\Phi_B/dT$), which resulted noticeably improved with respect to thermally grown $SiO_2$ layers.

These preliminary results suggest that reactive ion sputtering can be a promising technique to synthesize novel gate dielectric for wide band gap semiconductor devices technology.

## ACKNOWLEDGEMETS

The authors would like to thank E. Schilirò and R. Lo Nigro for fruitful discussions.
This work was carried out in the framework of the ECSEL JU project WInSiC4AP - *Wide Band Gap Innovative SiC for Advanced Power* (Grant Agreement n. 737483).

## REFERENCES

[1] M. Saggio, A. Guarnera, E. Zanetti,S. Rascunà, A. Frazzetto, D. Salinas, F. Giannazzo, P. Fiorenza, F. Roccaforte, Mater. Sci. Forum. 821-823, 660 (2015)

[2] F. Roccaforte, P. Fiorenza, G. Greco, R. Lo Nigro, F. Giannazzo, A. Patti, and M. Saggio, Phys. Status Solidi A, 211, 2063 (2014)

[3] B. J. Baliga, Silicon Carbide Power Devices, World Scientific Publishing Co. Pte. Ltd. , Singapore, 2005.

[4] M. Avice, U.Grossner, I. Pintilie, G. Svesson, M. Servidori, R. Nipoti, O. Nilsen, H. Fjellvag, J. Appl. Phys., 102, 054513, (2007)

[5] M. Usaman, C. Henkel, A. Hallen, ECS J.Solid State Sci.Techn. 2 N3087-N3091 (2013)

[6] M Vivona, G Greco, F Giannazzo, R Lo Nigro, S Rascunà, M Saggio, F Roccaforte, Semiconductor Science and Technology 29, 075018 (2014)

[7] E. Schilirò, R. Lo Nigro, P. Fiorenza,and F. Roccaforte, AIP ADVANCES 6, 075021 (2016)

[8] P. Fiorenza, V. Raineri, Appl. Phys. Lett. 88, 212112 (2006)

[9] P. Fiorenza, M. Vivona, F. Iucolano, A. Severino, S. Lorenti, G. Nicotra, C. Bongiorno, F.Giannazzo, F. Roccaforte Mater. Sci. Semicon. Processing 78, 38–42 (2018)

[10] M. Sometani, D. Okamoto, S. Harada, H. Ishimori, S. Takasu, T. Hatakeyama, M. Takei, Y. Yonezawa, K. Fukuda, H.- Okumura, J. Appl. Phys. 117 (2015) 024505-1–024505-6.

[11] P. Samanta, K.C. Mandal, , J. Appl. Phys. 121 (2017) 034501-1–034501-13.